# Artificial ageing of thin films of the indium-free transparent conducting oxide SrVO$_3$

M. Rath[1], M. Mezhoud[1], O. El Khaloufi[1], O. Lebedev[1], J. Cardin[2], Ch. Labbé[2], F. Gourbilleau[2], V. Polewczyk[3], G. Vinai[3], P. Torelli[3], A. Fouchet[1], A. David[1], W. Prellier[1], and U. Lüders[1]

[1]*Normandie Univ, ENSICAEN, UNICAEN, CNRS, CRISMAT, 6, boulevard du Maréchal Juin, F-14050 Caen, France.*

[2]*CIMAP, CNRS, ENSICAEN, UNICAEN, Normandie Univ, 6, boulevard du Maréchal Juin, F-14050 Caen, France.*

[3] *Istituto Officina dei Materiali (IOM)–CNR, Laboratorio TASC, Area Science Park, S.S.14, km 163.5, I-34149 Trieste, Italy*

## Abstract

SrVO$_3$ (SVO) is a prospective candidate to replace the conventional indium-tin-oxide (ITO) among the new transparent conducting oxide (TCO) generation. In this study, the structural, electrical, and optical properties of SVO thin films, both epitaxial and polycrystalline, are determined during and after heat treatments in the 150-250 °C range and under ambient environment in order to explore the chemical stability of this material. The use of these relatively low temperatures speeds up the natural ageing of the films, and allows to follow the evolution of their related properties. The combination of techniques rather sensitive to the film surface and of techniques sampling the film volume will emphasize the presence of a surface oxidation evolving in time at low annealing temperatures, whereas the perovskite phase is destroyed throughout the film for treatments above 200 °C. The present study is designed to understand the thermal degradation and long-term stability issues of vanadate-based TCOs, and to identify technologically viable solutions for the application of this group as new TCOs.

## Introduction

In recent years, SrVO$_3$ (SVO) thin films have received much attention as a promising material for transparent conducting oxides (TCOs)[1–3], used in optoelectronic devices and photovoltaic applications for instance[4,5]. The most attractive features of this material are a metallic conduction behavior with high conductivities[6–8], a high carrier density[9,10] in the range of $10^{22}$ cm$^{-3}$, a very weak work function[11] and a high visible-range optical transparency[2,7]. Moreover, avoiding the rare element Indium, the vanadium-based perovskite SVO is considered as a potential substitution for the standard TCOs like indium-tin oxide (ITO), F-doped $SnO_2$ (FTO) and Al doped ZnO.[5,12]

However, due to the presence of the 4+ oxidation state of the vanadium ion, which corresponds to one remaining electron in the ion's 3d orbital and is a rather unstable oxidation state, the material faces a specific risk of chemical degradation under ambient conditions, compromising its long-term stability. Losing the remaining 3d electron is energetically favorable for the V ion, so that SVO is prone to oxidation and to transformation to $Sr_2V_2O_7$ [13–15] or $Sr_3V_2O_8$ [8,16,17]. Despite being transparent phases, the related $V^{5+}$ oxidation state may increase substantially the electrical resistivity of SVO, and therefore deteriorates its functional properties as a TCO. The formation of these phases has been shown in the case of high temperature annealing under oxygen[13,16], but they may be also responsible for certain electrical contact issues observed with samples not having undergone special treatments, but which have only been stored at ambient conditions for some months. Due to the impact on the electrical conduction of SVO, the knowledge of the oxidation mechanisms at low to intermediate temperatures is important to avoid chemical instabilities of the Indium-free TCO in future applications.

Here, the objective of this study is to investigate the effects of heat treatments between 150 °C and 250 °C on the structural, chemical, electronic and optical properties of Pulsed Laser Deposition (PLD) grown epitaxial or polycrystalline SVO thin films. The intention is to accelerate the natural ageing effect in ambient atmosphere, without providing enough energy for long-range atomic movements in the film. In this way, only the effects being energetically favorable under ambient conditions will be observed, but the time scale will be shortened to allow for in-depth investigations. We have used both epitaxial and polycrystalline films in order to determine the effect of grain boundaries as possible preferential diffusion pathways on the ageing effects. Thus, the observed chemical modifications and their influence on the functional properties of SVO thin films will be a good indicator to determine the stability limits of this new TCO in possible applications. Our results show that the technologically interesting TCO phase of SVO is destroyed at heat treatments from 250 °C on, while for treatments at lower temperatures, the properties are modified, but the TCO character is preserved.

**Experimental Details**

For the SVO deposition by PLD, the target is composed of polycrystalline $Sr_2V_2O_7$ prepared by standard solid-state reaction. For the epitaxial growth of monocrystalline thin films, $SrTiO_3$ (001) monocrystalline substrates were used, whereas for the polycrystalline films, the growth was carried out on Eagle XG slim glass from Corning, using a 7 nm thick $TiO_2$ layer as a seed layer for the crystallization of the SVO layer. All substrates were cleaned before the introduction in to the deposition chamber with successive acetone and ethanol ultrasonic baths and dried with compressed air. The deposition was carried out in a PLD setup using an ultraviolet KrF excimer laser with a wavelength of 248 nm. During the deposition, the laser flux was maintained at 1.6-2.0 $J\ cm^{-2}$. The films were deposited under vacuum, with a pressure of about $1 \times 10^{-6}$ mbar when the substrate is at the deposition temperature. For the epitaxial films, the growth was carried out at 500 °C using a laser repetition rate of 3 Hz, while for the polycrystalline films, a growth temperature of 600 °C was used with a repetition rate of 5 Hz. The differences in the deposition parameters result from an independent optimization of the functional properties of the epitaxial and the polycrystalline films. The higher deposition temperature for the polycrystalline films is probably related to differences in the initial growth stages and the crystallization of SVO on a more heterogeneous surface. In both cases, the deposition was done with 5000 pulses, leading to films with a thickness of about 30 nm for the epitaxial films and about 25 nm for the polycrystalline films, as determined by X-Ray Reflectivity (XRR).

The thermal annealing of the various samples was carried out on a hot plate under ambient atmosphere, for 24 h. As deposited samples were located on the hot plate at room temperature, and the 4-point-probe for the resistivity measurement was positioned on the sample. Afterwards, the temperature of the hot plate was brought to the treatment temperature between 150 °C and 250 °C in some minutes. When the set point temperature was reached, a four-point-resistance head is brought into close contact to the sample surface and the resistivity measurement was started using a Keithley 2450 sourcemeter. The resistivity value shown at the time = 0 min corresponds to the room temperature resistivity of the sample before the beginning of the heat treatment. Every heat treatment was done on as deposited samples, avoiding subsequent heat treatments at different temperature. The heat treatments were done in a room without air conditioning, under a range of air humidity values around 60 %, typical for the Normandy climate. No influence of the air humidity was observed on the results.

For quality control purposes, the low temperature resistivity and Hall effect measurements were carried out using the van der Pauw contact geometry in a commercial Physical Properties Measurement System (PPMS) from QuantumDesign, under a maximum magnetic field of 9 T, with wirebonded contacts using a 25 µm Al wire.

The crystal orientation and crystalline quality of the thin films were examined by X-Ray Diffraction (XRD) for epitaxial films and Grazing Incidence X-Ray Diffraction (GIXRD) for polycrystalline ones with a Bruker D8 Discover or a PANalytical Materials Research diffractometer, both operating with monochromatic Cu $K_{\alpha 1}$ radiation ($\lambda$ = 1.5406 Å). The indexation of the $SrVO_3$ film reflections was done using the JCPDS file #06-9548. XRR measurements were carried out on the same machines. The XRR data was fitted using the software GenX[18].

The surface topography of the samples was analyzed by atomic force microscopy (AFM) (PicoScan AFM) using the tapping mode.

The optical properties of the samples were measured in the UV-visible-Near Infrared range using a Lambda 1050 Perkin-Elmer spectrophotometer operating in the conventional specular transmittance mode where light is transmitted through the sample and is collected by a detector aligned on the incident wave-vector axis. The experimental conditions including vertical beam mask, slit aperture, beam size as a 4 mm side length square, sample holder, the scan rate of 1 nm/sec, spectral range, angle of incidence, and polarization state were optimized and kept constant. The transmittance measurements of the materials were carried out in two steps. In the first step, a 100%T baseline spectrum was recorded by measuring with an empty setup (without sample). In the second step, the samples were placed in the sample holder. The whole measured transmittance spectra was corrected by the 100%T baseline (T=$I_{mes}/I_{100\%T}$). The transmittance spectra of the $SrTiO_3$ and the glass/$TiO_2$ substrates were measured with the same procedure. The presented transmittance spectra were then corrected by the substrate transmittance. In order to obtain realistic references for the substrate optical properties, regarding for example growth induced oxygen vacancies, both substrates were submitted to the deposition procedure, including the heating to the deposition temperature under vacuum, but without the deposition of the SVO film. The optical properties of these samples were used as a substrate reference for the calculation of the optical transmittance.

Transmission electron microscopy (TEM), including high-angle annular dark-field scanning TEM (HAADF-STEM) imaging and energy dispersive x-ray (EDX) elemental mapping experiments in STEM mode were carried out on an aberration double-corrected cold FEG JEM ARM200F microscope operated at 200 kV and equipped with a CENTURIO EDX detector, ORIUS Gatan camera and Quantum GIF. TEM cross-section samples were prepared using the Focused Ion Beam (FIB) technique with a Helios 450s FIB/SEM (Thermo Fisher, USA) instrument.

At the NFFA APE-HE beamline at Elettra synchrotron radiation facility[19], X-Ray Absorption Spectroscopy (XAS) measurements at the V $L_{2,3}$ edges were taken in linear horizontal polarization at room temperature, with the sample surface at 45° with respect to the impinging beam, in total electron yield (TEY) mode, for a probing depth of about ~ 5 nm. The intensity of the sample current was normalized to the incident photon flux current. For the qualitative analysis of the obtained XAS results, reference data of the three possible V oxidation states ($V^{3+}$, $V^{4+}$ and $V^{5+}$) were extracted from reference [20], and a linear combination of the three signals were fitted to the experimental data. While no background subtraction were performed, we weighted the different oxidization state signals by their cross section values. In our fit, no signature of the $V^{3+}$ contribution was detected in our different measurements, so that in a second step the data was fitted only with the $V^{4+}$ and $V^{5+}$ contribution. The prefactors of each contribution were then interpreted as the relative intensity.

## Results and Discussion

### Ageing effects on the film volume

Figure 1 shows the specular θ–2θ XRD profiles of the as deposited and the aged SVO thin films (150 °C, 200 °C and 250 °C) grown on a STO (001) substrate. The XRD spectra of the as deposited films have been collected during 48h following the deposition.

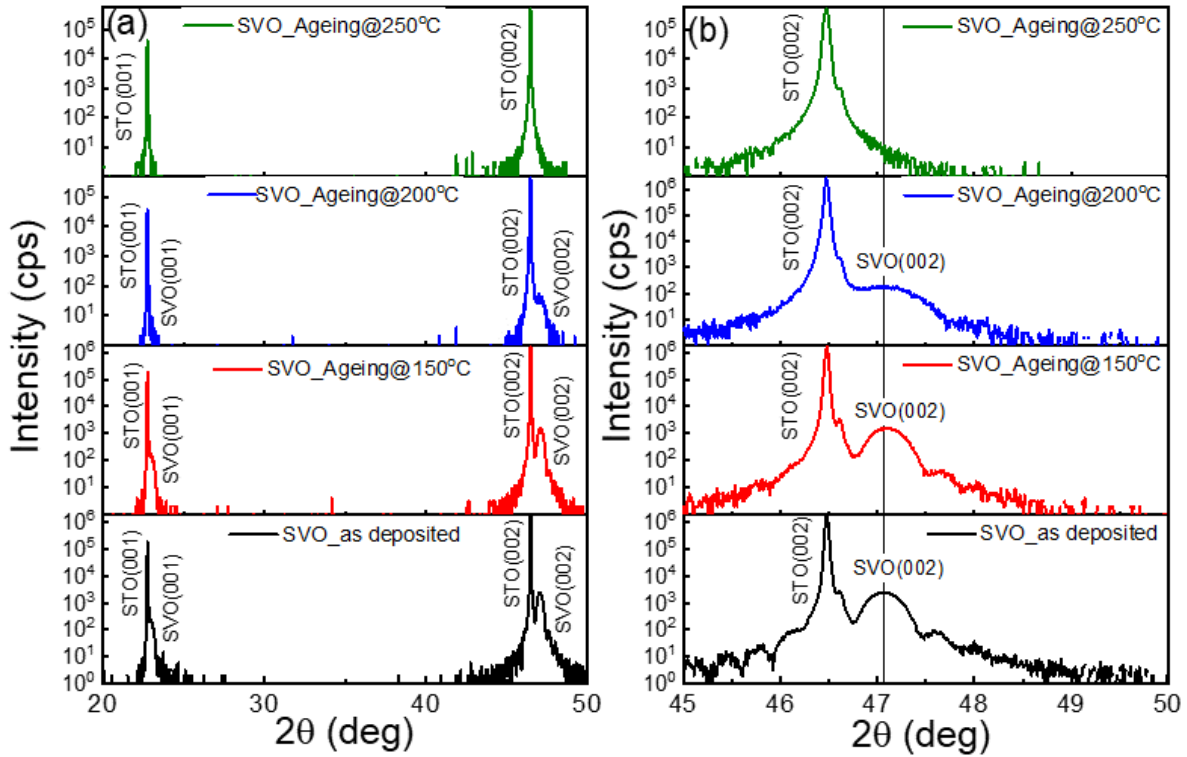

Figure 1 (a) Specular XRD results of epitaxial SrVO₃ films grown on SrTiO₃ (001), as deposited and after different heat treatments. (b) shows a zoom on the (002) peak of the same samples.

On the as deposited sample, the presence of solely the (001) and (002) film peaks of the cubic perovskite structure reflects the monocrystalline character of the film, oriented along the c-axis normal to the STO substrate surface. No traces of secondary phases are observed, and the presence of interference fringes at the SVO (002) peak indicates a high crystalline quality of the film with low roughness. A lattice parameter of 3.858 Å for the epitaxial films was calculated from the peak positions, comparable to what is obtained on other SVO films on SrTiO₃ (001) substrates[2,9,16,21–23]. Reciprocal space maps on the same system were published elsewhere[9] showing the coherently strained nature of the films.

After each thermal treatment under ambient conditions, the crystalline structure and the phase purity of the samples were examined by XRD (Figure 1). After an ageing at 150 °C for 24h, only a slight shift of the (002) SVO peak position is observed, while the intensity and the interference fringes are preserved. The peak shift corresponds to a small decrease of the lattice parameter to a value of 3.855 Å, probably related to the filling of oxygen vacancies present in the as-deposited film. The observed changes are more important after the heat treatment at 200 °C, where the (002) peak intensity clearly decreases, with a larger width and the disappearance of the interference fringes. All diffraction peaks related to the film disappear after a treatment at 250 °C of 24 h. As a result, the selected temperature range for the heat treatments includes a wide range of structural modifications of the epitaxial SVO thin films: a treatment at 150 °C causes only minor structural changes, but treatments at 200 and 250 °C destroy increasingly the perovskite structure. The sample shows no film-related diffraction peak following heat treatment at 250 °C, indicating the absence of long range order in the film.

Concerning the polycrystalline samples, Figure 2 shows the GIXRD diffractograms of the as deposited and the annealed films. The as deposited SVO film shows the main reflections of the cubic perovskite phase, without any secondary phase. Due to the low thickness (7 nm) of the $TiO_2$, no peaks are observed of this layer. When heated for 24 h, the perovskite structure of the films is preserved up to an annealing temperature of 200 °C, keeping its diffraction peak intensities unchanged. This indicates a stronger resistance to heat treatments at low temperature compared to the monocrystalline films. However, at 250 °C, again the characteristic diffraction peaks disappear, without the evolution of peaks related to a different phase.

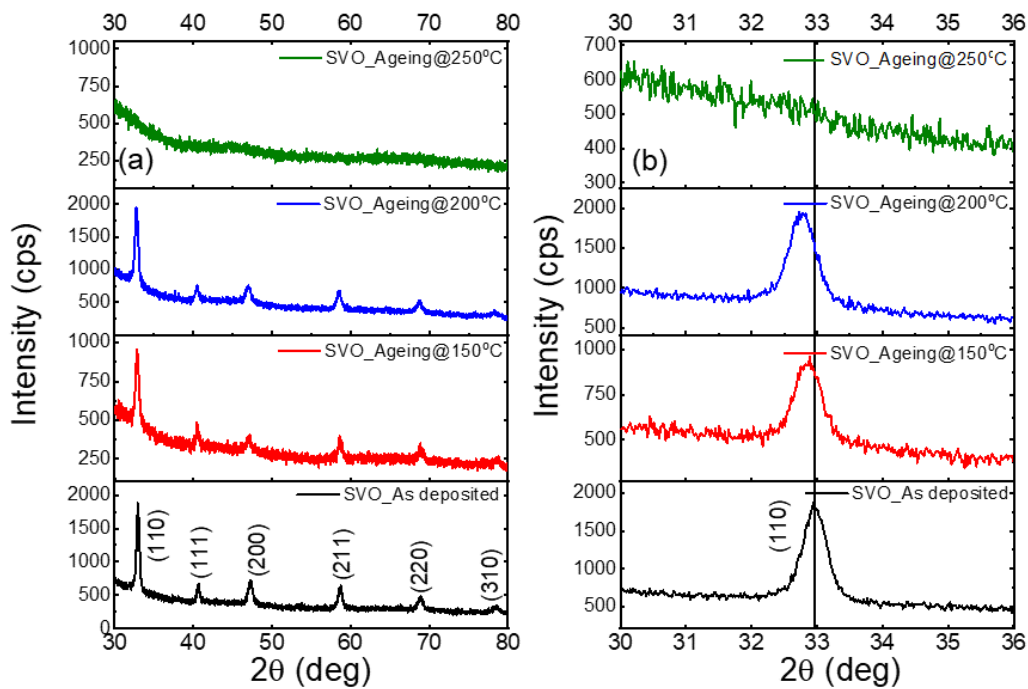

*Figure 2(a) GIXRD of polycrystalline SVO thin films on glass in the as deposited state and after ageing for 24h at different temperatures in air. (b) shows a zoom on the (110) peak of the same samples*

In Figure 2 (b), a close up of the most intense (110) peaks are shown. After annealing of the films at 150 and 200 °C, the peak position changes slightly. Using the (002) peaks, the lattice parameter of the SVO films was calculated to be 3.839, 3.854 and 3.864 Å for the as-deposited, 150 and 200 °C aged samples, respectively. In the case of the polycrystalline samples, the lattice parameter thus increases during heating, while an opposite behavior was observed for the monocrystalline films. The strain exerted by the $SrTiO_3$ substrate, absent in the polycrystalline films, may be the origin of this difference, but changes in oxidation due to the modification of the nucleation processes in the early growth stages or the existence of grain boundaries could also play an important role[24].

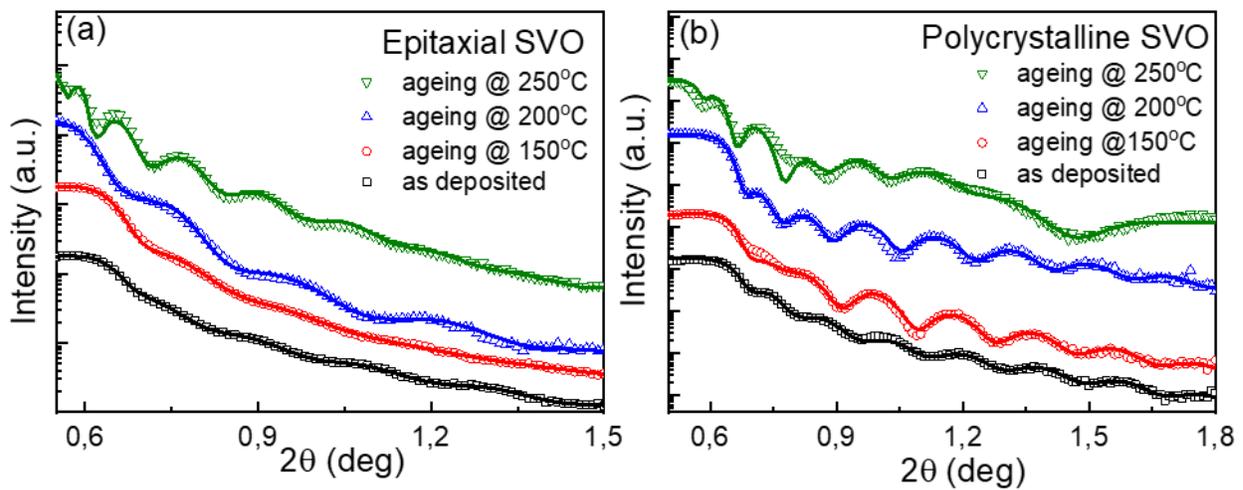

*Figure 3 XRR patterns of monocrystalline (a) and polycrystalline (b) SVO films, in the as deposited state and after the different heat treatments for 24 h. The experimental data is indicated symbols, and the fits are indicated as lines*

Figure 3(a) shows the XRR results of the epitaxial films, and the corresponding fits. The curve of the as deposited SVO film shows thickness oscillations, which are relatively small in amplitude due to the comparable electron density and therefore refraction index of $SrVO_3$ and $SrTiO_3$. The RMS roughness of the as-deposited film was determined to be below 0.4 nm by Atomic Force Microscopy measurements, indicating the weak oscillations are not due to an excessive roughness of the epitaxial film. The thickness of the as deposited film is determined to be 32(2) nm, which is consistent with previous calibration deposition rates. When the film is heated at 150 and 200 °C, a bilayer model was necessary for the fitting, using a bottom SVO layer with specifications corresponding to bulk SVO and a top layer, where the density was allowed to vary. From this analysis, an increasing thickness of the top SVO layer was observed: for an ageing at 150 °C, 8 nm at the top showed a decreased density for an overall film thickness of 34(2) nm, while the film aged at 200 °C shows a 10 nm thick top SVO layer and an overall film thickness of 36(2) nm. However, for the film heated at 250 °C, the full SVO film showed a decreased density reaching about 72% and a strong thickness increase, with a value going up to 52(3) nm. Before the heat treatment, the film has the same thickness as the as deposited one. The slope of the XRR curve does not change significantly, indicating that a possible roughness increase of the films stays limited. From AFM images, a RMS roughness increase from 0.4 nm to 2.5 nm is observed, consistent with the evolution of the XRR signal. This thickness increase, calculated to be 62% of the initial film thickness, is probably related to a massive oxygen introduction in the structure and a corresponding density change.

The thickness of the polycrystalline films was also characterized by XRR, the patterns and the fits are shown in Figure 3 (b). Here, for the as deposited sample, a bilayer model with one layer corresponding to the $TiO_2$ and another for SVO was used. The $TiO_2$ layer was found to have a thickness of about 10(2) nm, but it is complicated to determine a reliable value due to the large period of the corresponding oscillation. The thickness of the as-deposited polycrystalline film was calculated to be 28(2) nm, smaller than for the as-deposited epitaxial film. This may be due to the change in growth mechanism and a stronger desorption of the deposited material from the surface in the early stages of growth due to the absence of a well-ordered surface. Again, for the aged films, it was necessary to add a top SVO layer with varying density. Here, the film aged at 150 °C, showed a total thickness of 36 nm with a low density top layer of 7 nm. The film aged at 200 °C has a total thickness of 30 nm with a top SVO layer of also 7 nm. The polycrystalline film treated at 250 °C shows again a large thickness increase with a total thickness of 45 nm, corresponding to an increase of about 60 % of the initial film thickness.

The surface topography of all films were investigated by AFM scans in tapping mode, and the results of the polycrystalline films are shown in Figure 4. The films show a granular structure, with grain sizes of about 20 nm related to the polycrystalline structure of the films. The RMS roughness is about 0.88 nm in the as deposited case. For the aged films, the same granular structure is observed for the film aged at 150 °C (not shown here) and 200 °C, but the RMS roughness increases to 1.73 nm. For the film aged at 250 °C, the topography changes: the initial granular structure is still visible with grains in the 20 nm range, but an additional, larger surface modulation is observed (inset of Figure 4(c)) with rectangular structures of approximately 100 x 200 nm. The formation of embedded $Sr_3V_2O_8$ nanostructures were reported on the surface of epitaxial SVO films after a heat treatment under oxygen at higher temperatures[8,16]. The observed modulation of the surface for the film treated at 250 °C may be some reminiscent mechanism, less well defined due to the polycrystalline character of the film.

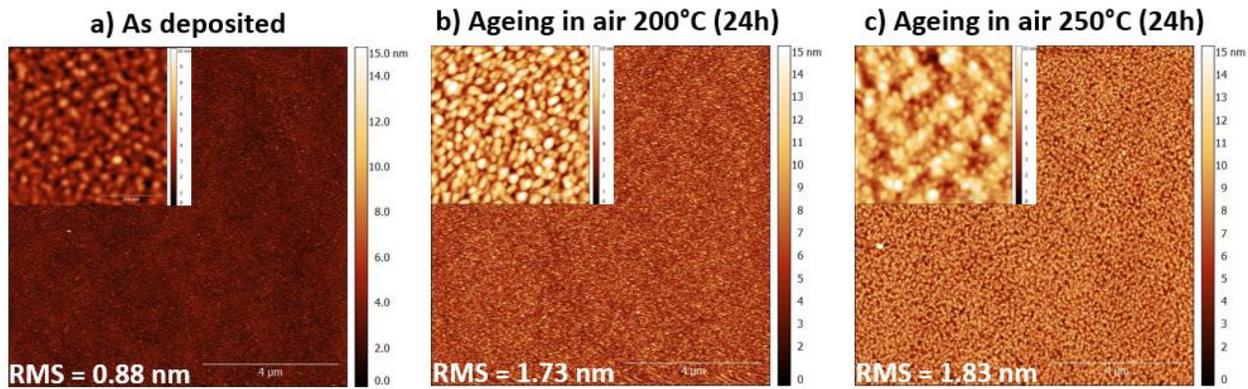

*Figure 4 Topography studies in the polycrystalline SVO films on glass, in the as deposited state (a) and after ageing treatments at 200 °C (b) and 250 °C (c). The size of the image is 10 x 10 μm², with a z-scale of 15 nm. The insets show close-ups of 1 x 1 μm² with a z-scale of 10 nm.*

The low temperature resistivity of the as-deposited films and Hall effect has been measured in order to control the electrical properties of both the epitaxial and the polycrystalline SVO films. The data is shown in Figure 5, highlighting the metallic character of the films in the as deposited state. The resistivity of the epitaxial films show the expected $T^2$ behavior of SVO, with resistivity values below 100 μΩcm. From Hall effect measurements (not shown here), a charge density of around $3 \cdot 10^{22}$ cm$^{-3}$ and for the room temperature mobility of about 1.5 cm²/Vs has been calculated, again comparable to what is observed in other thin films on $SrTiO_3$ [2,9,16]. The charge density is slightly enhanced compared to the theoretical value for a $3d^1$ configuration of the $V^{4+}$ oxidation state in SVO ($1.77 \cdot 10^{22}$ cm$^{-3}$), indicating the presence of some oxygen vacancies in the films related to the necessary deposition under vacuum. However, these vacancies were shown to be healed under an annealing of 100 °C[9], so that they do not play a role in the ageing treatments used in this study.

The temperature dependence of the resistivity of the polycrystalline samples (Figure 5) is less characteristic of SVO due to the presence of grain boundaries. Interestingly, the as deposited sample still shows a metallic temperature dependence, indicating that the grain boundaries do not perturb strongly the electrical conduction, as was also observed in optical measurements[3]. The influence of the grain boundaries on the electrical transport seems to be also limited, as the charge density is observed to be $1.7 \cdot 10^{22}$ cm$^{-3}$. This value, very close to the theoretical value, may indicate a lesser influence of oxygen vacancies than in the epitaxial films, but the chemically and structurally perturbed grain boundary regions are probably trapping some charges. The influence of the grain boundaries is more clearly visible in the charge mobility, where a value of 0.5 cm²/Vs is extracted from the Hall effect measurements, a factor of three lower than for the epitaxial films.

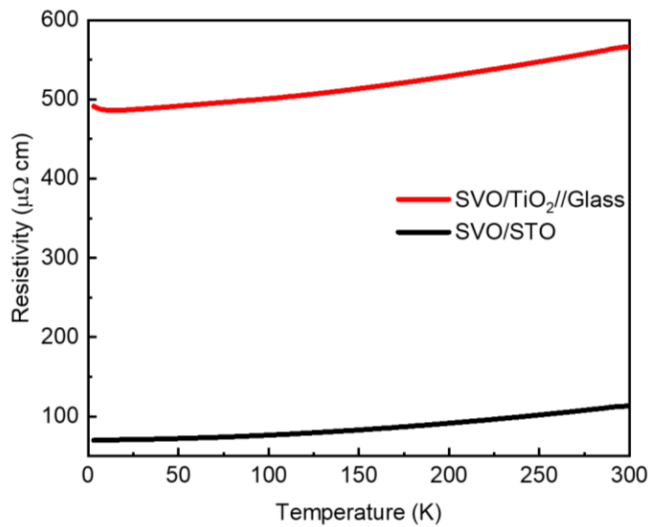

*Figure 5 Resistivity vs temperature of the as deposited films, both the epitaxial (black line) and the polycristalline one (red line)*

In Figure 6, the evolution of the resistivity of the SVO films during the heat treatment is shown for the different ageing temperatures, as well as for a sample kept at room temperature during the same period as a reference. Figure 6 (a) shows the evolution for the monocrystalline films grown on SrTiO$_3$ (001). The influence of the temperature treatment can be finely observed in these curves: while the resistivity of the sample kept at room temperature does not change, the samples kept at higher temperatures show an enhancement of the resistivity. For the samples aged at 150 and 200 °C, this enhancement is progressive and corresponds to a factor of 1.3 and 2.8 after 24h ageing. The sample treated at 250 °C also shows a progressive enhancement in the beginning, but after about 400 min, the evolution becomes stronger and the resistivity exceeds the measurement range after 600 min. Therefore, the destruction of the perovskite structure should intervene in this time scale.

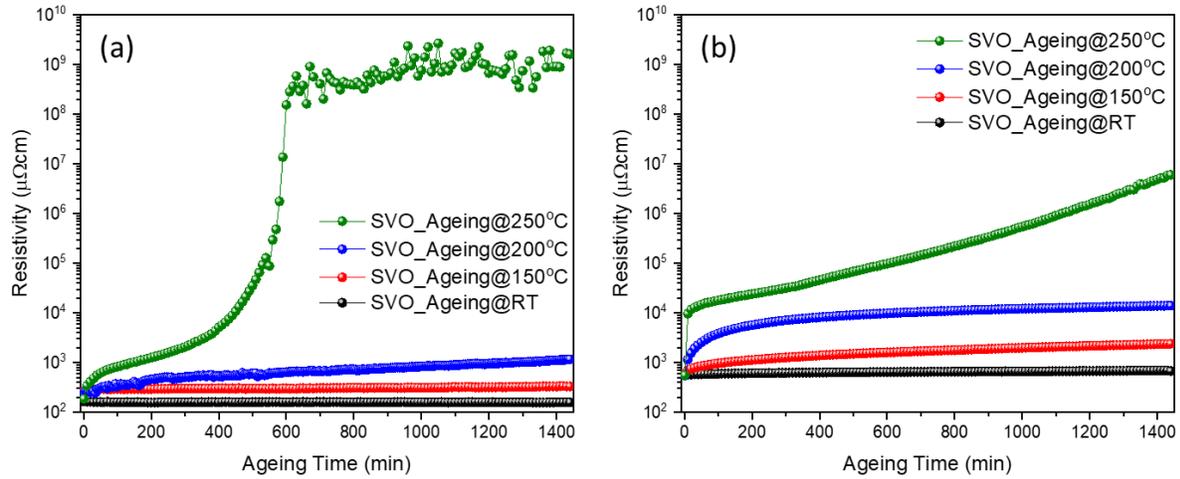

*Figure 6 Resistivity vs ageing time for the monocrystalline SVO films on SrTiO$_3$ (001) (a) and for the polycrystalline films on glass (b) for different ageing temperatures*

In case of the polycrystalline films (Figure 6 (b)), the same dependence on the ageing temperature can be observed, i.e. during the heat treatments, the resistivity increases progressively, while the sample kept at room temperature does not show such an enhancement. However, the effect at small temperatures, 150 and 200 °C, is stronger, where the resistivity enhancement reaches factors of 4.1 and 25.7 after 24h ageing, respectively. This change of behavior is probably related to the presence of the grain boundaries in these samples. The exact role of the grain boundaries in SVO is not known for the moment, but the chemically closely related VO$_2$ phase shows a complex interplay of fast grain boundary oxygen diffusion and the presence of oxygen vacancies under air heat treatments at 450 °C[25]. Comparable mechanisms are possible in SVO, too, leading to this complex behavior. The film aged at 250 °C shows the strongest enhancement of the resistivity, especially in the beginning of the heat treatment. Although this sample does not show an electric breakdown comparable to the monocrystalline film, maybe again because of the presence of grain boundaries, the slope of the resistivity evolution with time is clearly steeper than for lower ageing temperatures.

In order to probe the optical properties of SVO thin films after ageing, optical transmittance spectra of as deposited and aged epitaxial films (Figure 7 (a,b)) and polycrystalline films (Figure 7 (c,d)), were measured at room temperature in the visible range. For both kinds of films, epitaxial and polycrystalline, the transmittance decreases with the wavelength between 480 and 400 nm due to a characteristic interband transition of SVO from the low-lying O bands to the V valence band[3]. On the other hand, films heated to 250°C no longer show this decrease in transmittance which can be explained by the loss of the SVO perovskite structure as evidenced by XRD. In order to compare the transmittance T over the full visible range (λ = 400 – 800 nm), a normalized integrated transmittance factor $\eta_T$ was built using the ideal transmittance of 100 % as $T_{max}$:

$$\eta_T = 100 * \frac{\int_{\lambda_{min}}^{\lambda_{max}} T(\lambda) d\lambda}{\int_{\lambda_{min}}^{\lambda_{max}} T_{max}(\lambda) d\lambda}$$

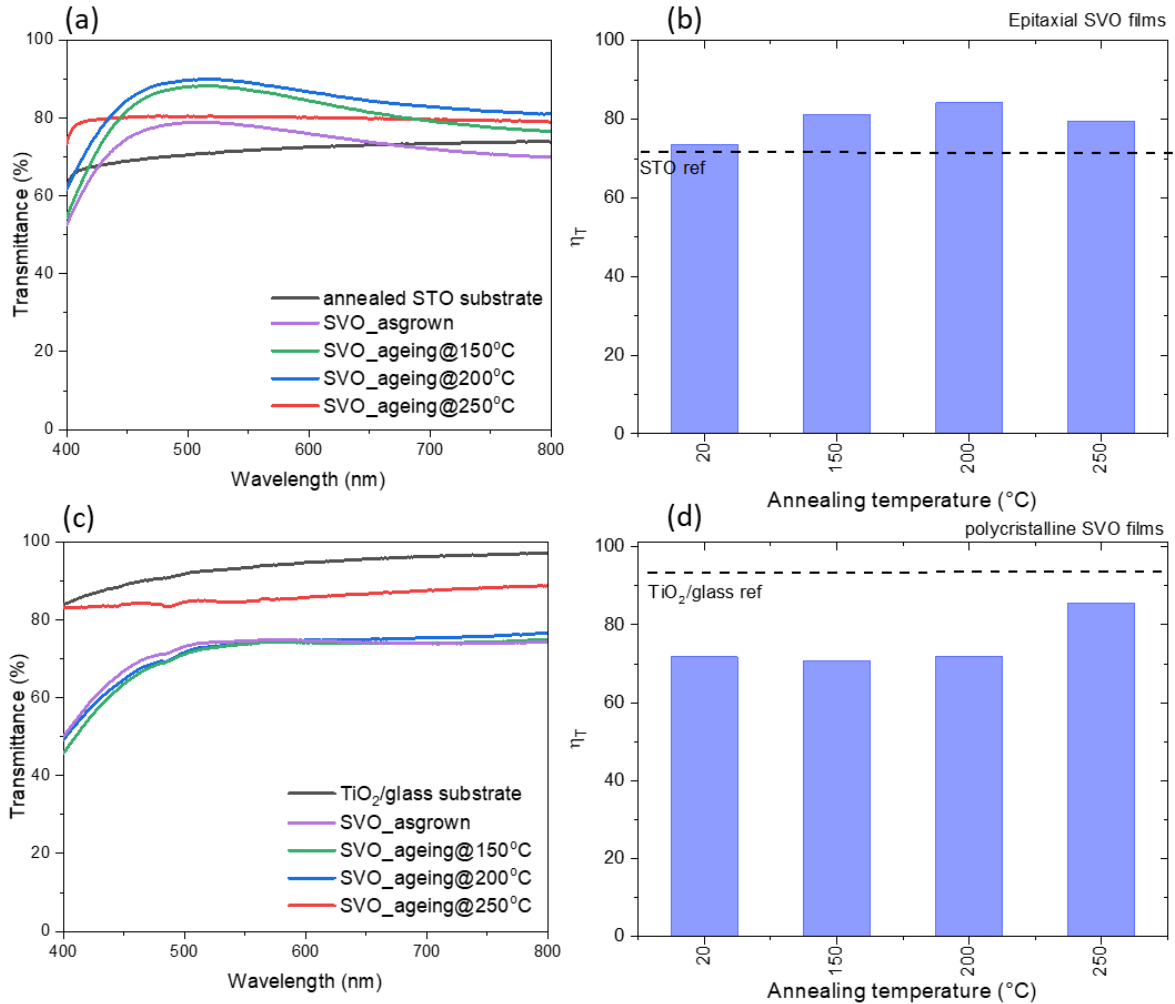

Figure 7 The variation of the optical transmittance (a,c) and the normalized transmittance factor $\eta_T$ (b,d) of epitaxial (a,b) and polycristalline (c,d) SVO thin films in the as deposited state and after 24 h of ageing at different temperatures.

For the epitaxial films (Figure 7 (a)), $\eta_T$ show a maximum for the ageing temperature of 200°C, related to its excellent transmission of about 90 % at 500 nm. For the polycrystalline films (Figure 7 (d)), $\eta_T$ slightly decreases with ageing up to 200°C and abruptly increases for an ageing at 250°C. This latter sample is known to have lost its SVO structure and might be therefore more transparent principally due to the absence of the interband transition, underlining the effect of the free charge carriers on the optical properties of SVO. In addition, $\eta_T$ shows that for ageing temperatures up to 200°C, the epitaxial films are more transparent than the polycrystalline films with a maximum difference of 12% for heating at 200°C. This difference is probably related to the presence of grain boundaries in the polycrystalline films, leading to supplementary optical extinction due to supplementary absorption and scattering.

Therefore, the ageing treatments have a significant effect on the structure, the topography, the electrical resistivity and the optical properties of the SVO films, both in the polycrystalline and in the epitaxial form. In both cases, the film properties are only slightly influenced after an ageing at 150 °C, but show a destruction of the SVO structure and its principal functional properties after an ageing at 250 °C. The samples heated at 200 °C show some kind of transitional behavior, where in the case of epitaxial films, the quality of crystalline structure is already strongly diminished, while the other properties are only influenced peripherally. In the case of the polycrystalline films, the structure is relatively preserved at this ageing temperature, but the electrical resistivity is already strongly enhanced.

**Ageing effects on the film surface**

In order to gain more insight into the structural, microstructural and chemical effects of the ageing on the surface of the SVO films, we have carried out in-depth characterization by TEM both on as deposited and on the 200 °C aged polycrystalline SVO films. The results on the 200 °C aged film are shown in Figure 8. The images show a typical columnar grain structure of the SVO film, where most of the grains start to crystallize at the interface with the $TiO_2$ buffer layer. The $TiO_2$ buffer layer has a thickness around 10 nm and can be clearly seen in HAADF-STEM image due to darker with respect to SVO film contrast. The grain structure of the $TiO_2$ buffer layer is less well visible, related to its film thickness, but also to a smaller grain size and overlapping effect. However, in the intermediate magnification image (Figure 8 (a)), grains can be identified also in this layer. The HAADF-STEM images also show a layer without evident long-range crystalline structure on the top of the SVO grain structure, with a strongly varying thickness and slightly darker contrast. However, the performed EDX-STEM elemental mapping confirmed an identical SVO film composition of this layer (Figure 8). Moreover, the images of the as deposited film (not shown here) have shown a roughness comparable to that of the AFM images, lower than the roughness of the crystalline phase of the aged film. Therefore, this top layer forms during the ageing treatment by a transformation of the different grains. The strong roughness of the interface between the crystalline grains and the top layer can be explained by a different transformation rate of the different crystalline orientations.

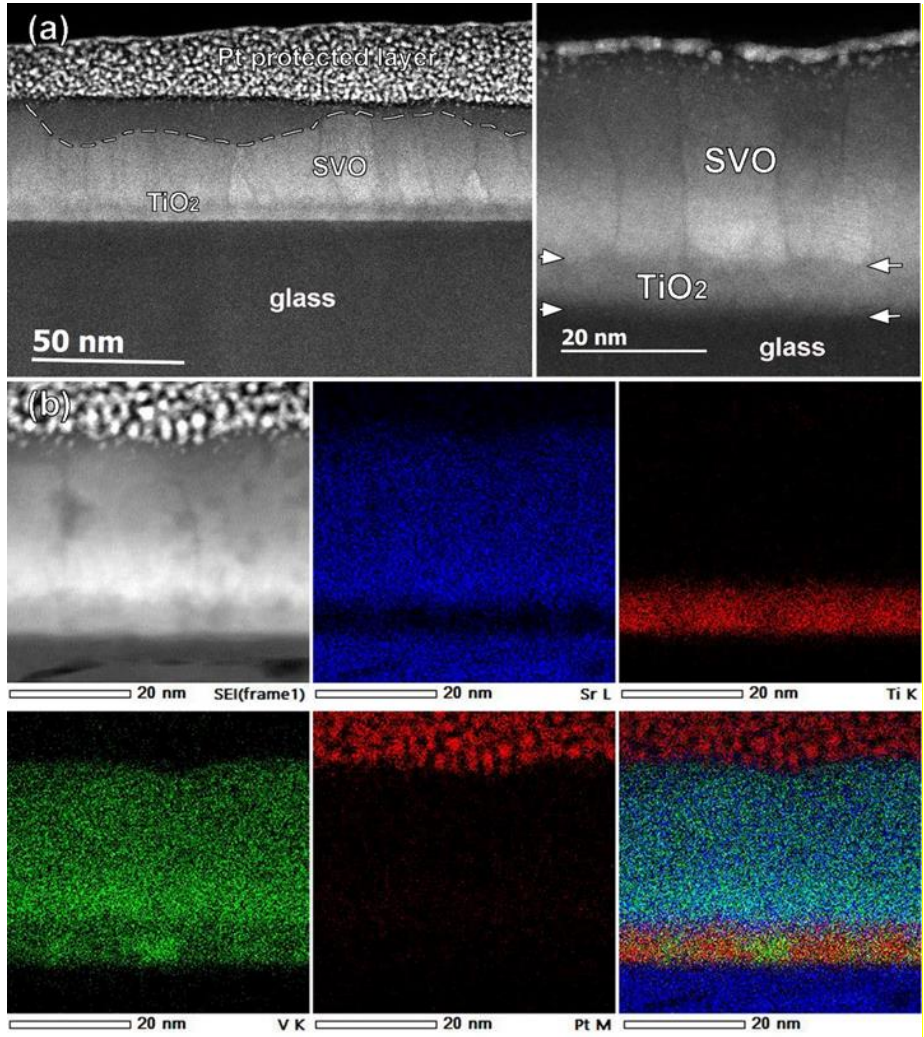

*Figure 8(a) Low and intermediate magnification HAADF-STEM images of a cross section of a polycrystalline SVO film aged at 200 °C for 24h and (b) simultaneously acquired EDX-STEM elemental mapping for Sr L, Ti K, V K, Pt M edges, and an overlayed color image. The dashed line in (a) depicts the appearance of a SVO amorphous layer at the surface of the SVO film. The buffer TiO$_2$ layer is marked by white arrows.*

When regarding the elemental structure of the film, the Sr and the V distribution seems to be homogeneous throughout the SVO films and the amorphous layer. A certain gradient of the Sr and V signal can be observed in the first 5 to 7 nm of the SVO film, but this is related to a thickness variation of the cross section sample. The Sr/V ratio through the film thickness is constant. No sign of surface segregation, for example of Sr, is observed. On the other hand, the interface with the TiO$_2$ buffer layer is chemically less well defined: in the overlayed image in the right bottom corner of Figure 8, the interdiffusion of V into the TiO$_2$ layer can be observed. This is also the case in the as deposited film (not shown here), where such an interdiffusion is also observed. Therefore, the interdiffusion occurs during the growth of the films, and is not a result of the ageing treatment. In the elemental maps, this badly defined interface between the buffer layer and the polycrystalline SVO films can be also observed: the Ti map (red) shows a gradual top interface, overlapping with the Sr signal (blue), and the V map (green) shows an important presence throughout the TiO$_2$ buffer layer, although less than in the SVO layer.

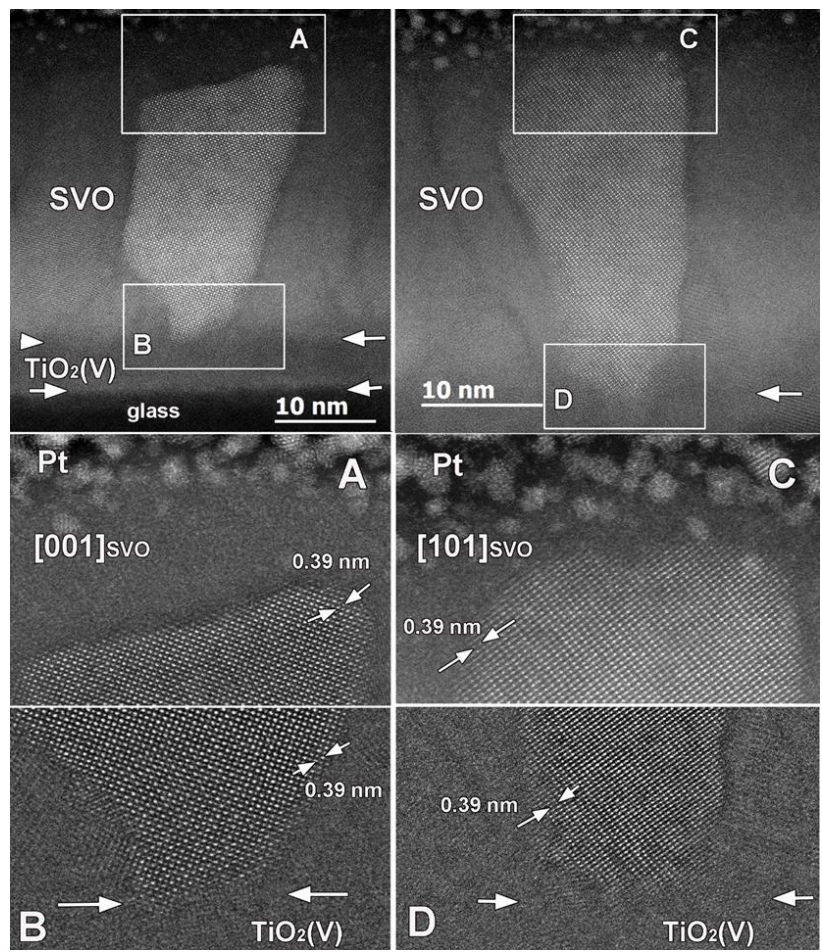

*Figure 9 Cross section high resolution HAADF-STEM image of a single SVO grain along [001] (left) and [101] (right) zone axis of SVO perovskite structure. A,B,C and D are magnified images of representative regions selected in the images above.*

This interdiffusion at the interface with the buffer layer does not seem to influence the SVO structure in the grains. First, the XRD patterns showed an undisturbed SVO phase for the as deposited films, indicating that the majority of the crystalline part of the film is actually $SrVO_3$. High magnification images of the 200 °C aged sample also confirm these findings (Figure 9): The interplanar distance of the SVO grains corresponds to 0.39 nm, which is consistent (inside the error bars) with the XRD results. Therefore, even in the aged films, the crystalline grains keep a crystalline structure comparable to the SVO pure phase, without indications of V vacancies (due to the interdiffusion, for example) or Sr segregation in the grains. In these images, it can also be observed that the interface between the grain and the amorphous top layer is very abrupt, and that no visible deterioration of the SVO phase precedes the top layer.

Despite no visible chemical changes in the top layer, the progressive disappearance of the perovskite structure with the annealing temperature and the evolution of the electronic properties in the aged samples and their thickness enhancement observed by XRR in the case of an ageing temperature of 250°C raise the question about the chemical state of these films and especially of the top layer. Therefore, we have carried out XAS measurements on the polycrystalline thin films in order to identify the V oxidation state in the samples aged at 200 °C. Since the probing depth is around 5 nm, these measurements are extremely sensitive to the first nanometers of the surface, i.e. the range of thickness where the top layer is expected to be majoritarian in the aged state. Measurements were taken in two steps. First, an ageing treatment at 200 °C was applied during 24 h to as deposited samples, and the XAS was measured. Then, the samples were cleaned with a 30 s bath in water, by agitating them manually. This water bath is known to dissolve higher Sr-V-oxides on a quicker time scale than the cubic $SrVO_3$[26]. Therefore, the surface of the washed sample will be representative of the film structure in absence of possible surface modifications due to the ageing treatments.

The XAS V $L_{2,3}$ edges of the aged surface of a 200 °C treated polycrystalline sample, i.e. the aged state, are shown in Figure 10, black curve, labelled "before water". The shape of the two edges, and especially the narrow V-$L_3$ edge and its pre-peak shoulder, is typical of a strong contribution of $V^{5+}$ [20], which is characteristic of the formation of $V_2O_5$, but also of $Sr_3V_2O_8$ or $Sr_2V_2O_7$, two phases that were observed at the surface of SVO films heated in air at higher temperatures[13,17,26]. For qualitative analysis, a linear combination fit using reference spectra extracted from reference [20] has been performed taking into account the weight of each oxidization state at these absorption energies (dashed lines in Figure 10). The results confirm the large majoritarian contribution of $V^{5+}$.

Therefore, although the elemental EDX-STEM maps show a similar composition, a precise characterization of the surface of the aged films undergoes a clear chemical change, principally related to an over-oxidation of the Vanadium ions.

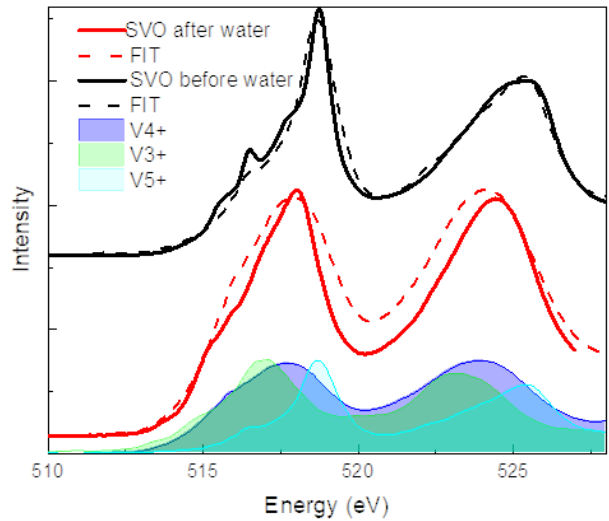

Figure 10 XAS measurements at V $L_{2,3}$ edges of a polycrystalline film aged at 200 °C, before (black) and after (red) the water treatment. The dashed lines are fits with reference lines from reference [23].

After the samples have been cleaned with liquid water, the XAS spectra changed substantially. The spectral shape of the V edges indicates now a majority $V^{4+}$ state, showing a complete change in the shape of the spectrum, especially concerning the relative intensities between the two edges, in agreement with what reported for other $V^{4+}$ compounds[20,27,28]. A comparison with bulk $SrVO_3$ from reference [27] or our fits gives a good qualitative agreement. The modification of the XAS spectra during the process clearly indicates how the amorphous top layer has a 5+ character, while the clean $SVO_3$ layer below it a 4+ one, which is the oxidation state expected for stoichiometric SVO.

In order to link the surface amorphous layer observed by TEM with the water-soluble $V^{5+}$ rich phase observed by XAS, we have carried out XRR studies on the 200 °C aged polycrystalline sample before and after washing with water (Figure 11 (a)). The thickness of the seed layer and the SVO film before the water etching, i.e. in the aged state, is 39(2) nm, which reduces to 33(2) nm after the water etching. Therefore, the thickness of the water-soluble layer is about 6 nm (as shown in Figure 11 (c), the seed layer is conserved through the water etching). This thickness is in agreement with what obtained with XAS measurements (Figure 9), taking into account the rough morphology of the surface after ageing process measured via AFM (Figure 4). In Figure 11 (b), the influence of the water etching on the crystalline structure is shown, illustrated by a close-up of the GIXRD (110) SVO reflection. No significant change of the peak is observed, indicating that the water etching does not act on the SVO phase, but only the surface layer with the $V^{5+}$ component is etched.

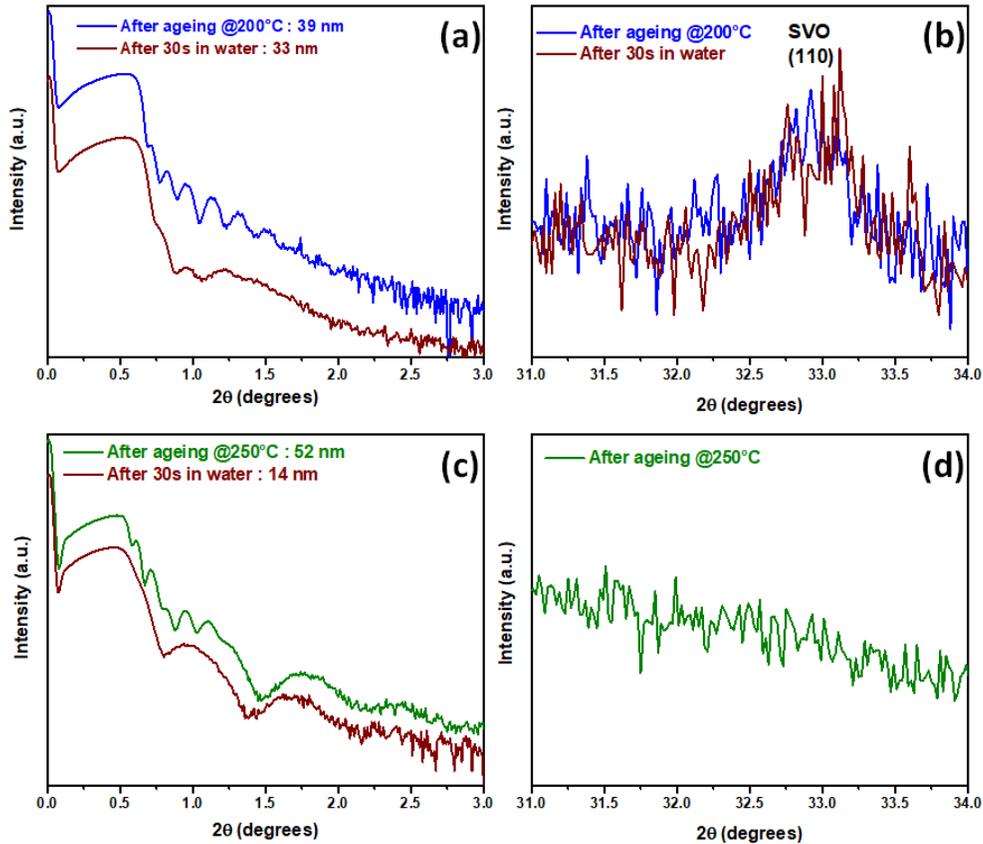

Figure 11 XRR (a, c) and GIXRD (b, d) on the (110) SVO reflection for a polycrystalline SVO film on glass aged at 200 °C (a and b) and at 250 °C (c and d), before and after the water etching treatment

The same kind of experiment was also done on a polycrystalline SVO film aged at 250 °C for 24 h. For this film, the GIXRD measurements did not show any reflection related to the SVO phase (Figure 2), so that it is probable that at such a high ageing temperature, the $V^{5+}$ phase extends through the whole layer. The thickness of the aged film before the water treatment determined by XRR measurements (Figure 11 (c)) is 52 nm, including the thickness of the $TiO_2$ seed layer. When the sample is washed in water, the short period oscillation disappears, and the only oscillation period still observed is the one related to the $TiO_2$ seed layer. Therefore, the water etching dissolves most of the SVO film aged at 250 °C, indicating a full $V^{5+}$ state, and not only on the surface as observed for the ageing treatment at 200 °C.

**Conclusions**

The effect of low temperature heat treatments in air on the chemical, structural, electrical and optical properties of SrVO$_3$ films, both in the monocrystalline as in a polycrystalline state, have been studied. While for ageing temperatures at 150 °C and at 200 °C, the SVO phase is preserved under a V$^{5+}$ rich surface layer, an ageing at 250 °C destroys the technologically interesting TCO phase. These results are important to assess the stability of SVO, especially for its applications as a new Indium-free TCO. From the here reported results, different conclusions can be drawn. First, under ambient conditions, SVO films and their properties should be rather stable, as a possible oxidation of the surface concerns only the extreme surface of the layer, so that the chemical modifications do not alter the functional properties throughout the SVO layer even after extended expositions to air. Second, temperature treatments or a working temperature up to 150 - 200 °C in oxygen containing atmospheres are supported by the SVO layers, although a certain increase of the electrical resistivity compared to the pristine layers may be observed. Prolonged exposition to ambient atmospheres at temperatures exceeding 200 °C should be strictly avoided, as this may lead to the destruction of the SVO phase, and therefore to a strong resistivity increase, although the films stay transparent. And third, a repetitive exposition to water should also be avoided, as relatively short water treatments lead to the dissolution of the passivating surface layer. If between different water expositions, the time is long enough for the surface layer to form, the thickness of the SVO layer will decrease slightly after each water contact.

These stability limits of SVO may seem to be rather severe compared to other oxides, where the transition metal ion has a more stable oxidation state. Also, the presence of Sr seems to be related to the water solubility of vanadates[29], which is interesting for the restitution of the SrVO$_3$ surface after ageing or sacrificial layer applications, but nor for the use of SVO as a TCO under humid conditions. But two simple pathways to extend the stability limits of SVO should be studied. First, a protection layer to be deposited on top of the SVO film can be developed. For SVO, TiO$_2$ was tested[30], but a comparable approach as what has been done for SrTiO$_3$ would be possible, too. Here, Al-based oxides or even an ultra-thin metallic Al film, oxidized by the SrTiO$_3$ itself, have shown a protecting effect concerning oxygen exchange with the atmosphere, even at high temperatures[31]. Second, the ageing is related to the presence of Sr and the unstable oxidation V$^{4+}$ state. If one or both of these critical ingredients for the ageing could be avoided in perovskite TCOs, ageing effects may be strongly diminished. SrVO$_3$ is just one representative of this new class of materials, other compounds as CaVO$_3$ [1], SrNbO$_3$ [32] or SrMoO$_3$ [33] have comparable functional properties and may be chemically more stable. But even more, Ca based compounds as CaNbO$_3$, CaMoO$_3$ or the Ca-Al-O system may be interesting alternative candidates, being less subject to the observed ageing effects in SVO. These materials can be synthesized and do show metallic conductivity[33,34], but their ageing has not yet been studied.

**Acknowledgements**

The authors would like to acknowledge financial support from the CNRS prematuration project CoCOT, and the Normandie region through the projects RIN PLDsurf and Cibox, and the funding of the PhD thesis of O.E.K. This work has been partially performed in the framework of the Nanoscience Foundry and Fine Analysis (NFFA-MUR Italy Progetti Internazionali) facility and the CNRS Federation IRMA - FR 3095.